\documentclass[aps,prl,twocolumn,showpacs]{revtex4}
\usepackage{graphics}
\usepackage{amsmath}
\usepackage{bm}

\begin{document}

\title{Conductance of a quantum wire in the Wigner crystal regime}

\author{K. A. Matveev} 

\affiliation{Department of Physics, Duke University, Box 90305, Durham, NC
  27708}

\date{October 10, 2003}

\begin{abstract}
  We study the effect of Coulomb interactions on the conductance of a
  single-mode quantum wire connecting two bulk leads.  When the density of
  electrons in the wire is very low, they arrange in a finite-length
  Wigner crystal.  In this regime the electron spins form an
  antiferromagnetic Heisenberg chain with exponentially small coupling
  $J$. An electric current in the wire perturbs the spin chain and gives
  rise to a temperature-dependent contribution of the spin subsystem to
  the resistance.  At low temperature $T\ll J$ this effect is small, and
  the conductance of the wire remains close to $2e^2/h$.  At $T\gg J$ the
  spin effect reduces the conductance to $e^2/h$.
\end{abstract}


\maketitle

Experiments with short one-dimensional (1D) conductors (quantum wires)
have demonstrated \cite{old} that their conductance is quantized in units
of $2e^2/h$.  The universality of this result is readily understood in the
model of non-interacting electrons.  In this approach the conductance is
given by $G=e^2 \nu v_F$, where $v_F$ is the Fermi velocity of electrons
and $\nu=2/hv_F$ is their density of states in one dimension.  It is
remarkable, however, that in most experiments observing the quantization
of conductance the Coulomb interactions between electrons are not weak,
$(na_B)^{-1}\gtrsim1$.  (Here $n$ is the electron density,
$a_B=\varepsilon\hbar^2/me^2$ is the Bohr radius, $\varepsilon$ is the
dielectric constant, and $m$ is the electron effective mass.)
Electron-electron interaction in a 1D system is expected to lead to the
formation of a Luttinger liquid with properties very different from those
of the non-interacting Fermi gas.

The conductance of an infinite Luttinger liquid was studied by Kane and
Fisher \cite{kane}, who found that it does depend on the interactions.
In particular, in the case of repulsive interactions the conductance is
below the universal value of $2e^2/h$.  The discrepancy between the theory
\cite{kane} and experiments \cite{old} is usually attributed to the fact
that in order to measure conductance of a quantum wire, it has to be
connected to two-dimensional Fermi-liquid leads.  As electrons move from
the wire into the leads, the interactions between them are gradually
reduced to zero.  At low frequency $\omega\ll v_F/L$ the conductance of
such a system is dominated by the leads, and the universal value $2e^2/h$
is restored \cite{maslov}.

The key assumption leading to this result is the applicability of the
Luttinger liquid description of the interacting electron system in a
quantum wire.  In this paper we show that if the interactions are strong,
$(na_B)^{-1}\gg1$, the Luttinger liquid picture remains valid only at
exponentially low temperatures, and study the corrections to the quantized
conductance at higher temperatures.

The low-energy properties of an interacting 1D electron system are most
conveniently described by the bosonization approach.  In the case of weak
interactions the Hamiltonian of the system can be presented \cite{schulz}
as
\begin{equation}
  \label{eq:separation}
  H = H_\rho + H_\sigma,
\end{equation}
where
\begin{eqnarray}
  \label{eq:H_rho}
  H_\rho&=&\int\frac{\hbar u_\rho}{2\pi}\left[\pi^2 K_\rho \Pi_\rho^2 
                         +K_\rho^{-1}(\partial_x\phi_\rho)^2\right]dx,
\\
  \label{eq:H_sigma}
  H_\sigma&=&\int\frac{\hbar u_\sigma}{2\pi}
                         \left[\pi^2 K_\sigma \Pi_\sigma^2 
                         +K_\sigma^{-1}(\partial_x\phi_\sigma)^2\right]dx
\nonumber\\
          &&+\frac{2g_{1\perp}}{(2\pi\alpha)^2}\int 
             \cos\left[\sqrt8 \phi_\sigma(x)\right] dx.
\end{eqnarray}
Here the fields $\phi_\rho,\Pi_\rho$ and $\phi_\sigma,\Pi_\sigma$ describe
the excitations of the charge and spin modes, respectively, and satisfy
the bosonic commutation relations:
$[\phi_\nu(x),\Pi_{\nu'}(x')]=i\delta_{\nu\nu'}\delta(x-x')$.  Parameters
$K_\rho$, $K_\sigma$, and $g_{1\perp}$ are determined by the interactions
between electrons, $u_\rho$ and $u_\sigma$ are the velocities of
propagation of spin and charge excitations, and $\alpha$ is a
short-distance cutoff.

An important feature of the Hamiltonian (\ref{eq:separation}) is the
separation of the charge and spin variables \footnote{The Hamiltonian
  (\ref{eq:separation})--(\ref{eq:H_sigma}) may contain irrelevant
  perturbations of third and higher powers in bosonic fields that violate
  the spin-charge separation.  They can be neglected at temperatures small
  compared to the Fermi energy.}.  This property is preserved even when
bias is applied to the wire, because the electric potential couples only
to the charge density, and does not excite the spin modes.  As a result,
the relatively complicated form (\ref{eq:H_sigma}) of the Hamiltonian
$H_s$ is not expected to affect the conductance of the quantum wire.

The derivation \cite{schulz} of the bosonized Hamiltonian assumes weak
interactions between electrons.  On the other hand, the form
(\ref{eq:separation})--(\ref{eq:H_sigma}) of the Hamiltonian is universal,
i.e., with the appropriate choice of the parameters $K_{\rho,\sigma}$,
$u_{\rho,\sigma}$, and $g_{1\perp}$ it is expected to describe the
low-energy properties of 1D electron liquids with arbitrarily strong
interactions.  It will be instructive to obtain the Hamiltonian
(\ref{eq:separation})--(\ref{eq:H_sigma}) from the model of fermions with
strong Coulomb interactions, and to estimate the values of the parameters
entering Eqs.~(\ref{eq:H_rho}),~(\ref{eq:H_sigma}).

At low density $n\ll a_B^{-1}$ the potential energy of the electron
repulsion $\sim e^2n/\varepsilon$ is much larger than their kinetic energy
$\sim\hbar^2n^2/m$.  Thus to first approximation one can view the
electrons in a quantum wire as a Wigner crystal of particles repelling
each other with strong Coulomb forces.  Assuming that the Coulomb
interaction is screened at large distances by a metal gate parallel to the
wire, the density excitations of the Wigner crystal (plasmons) are
acoustic waves.  Thus the low-energy density excitations of this system
must be described by the Hamiltonian of the form (\ref{eq:H_rho}).  The
plasmon velocity in a one-dimensional Wigner crystal at a distance $d$
from the gate is $u_\rho=\sqrt{e^2n/mC}$, where $C=\varepsilon/[2\ln(\zeta
nd)]$ is the capacitance per unit length between the crystal and the gate
\cite{ruzin}.  (Here $\zeta$ is a numerical coefficient determined by the
geometry of the gate \footnote{In the case of a gate being an infinite
  conducting plate, $\zeta\approx 8.0$, Ref.~\onlinecite{ruzin}.}.)
Parameter $K_\rho$ can be found by comparing the second term in the
integrand of Eq.~(\ref{eq:H_rho}) with the charging energy per unit length
$E_C=e^2 \delta n^2/2C$, where $\delta n(x)$ is the deviation of the
density of electrons from its mean value.  The bosonization procedure used
in the Hamiltonian (\ref{eq:H_rho}) identifies $\delta n=
-\frac{\sqrt2}{\pi}\partial_x\phi_\rho$.  Comparing $E_C$ with the second
term in Eq.~(\ref{eq:H_rho}), one then finds
$K_\rho=(\pi\hbar/2)\sqrt{nC/me^2}$.  Substituting the above expression
for $C$, we summarize
\begin{equation}
  \label{eq:rho-parameters}
  u_\rho = \frac{v_F}{K_\rho}, 
  \quad 
  K_\rho=\frac{\pi}{2} \sqrt{\frac{na_B}{2\ln(\zeta nd)}},
\end{equation}
where the parameter $v_F\equiv\pi\hbar n/2m$ is defined as the Fermi
velocity in a gas noninteracting electrons at density $n$.  As expected, at
$(na_B)^{-1}\gg1$ strong Coulomb interactions result in $K_\rho\ll1$ and
$u_\rho\gg v_F$.

In the limit of strong coupling the energy of a Wigner crystal does not
depend on the spins of electrons.  Indeed, electrons localized near their
equilibrium positions can be viewed as distinguishable particles, and the
antisymmetrization of the wavefunctions does not affect the energies of
the eigenstates.  This picture is violated if one allows for the
possibility of overlap of the wavefunctions of neighboring electrons.  The
overlap is due to the possibility of tunneling through the potential
barrier $e^2/\varepsilon r$ separating the electrons.  To first order in
tunneling, the overlap occurs only between the nearest neighbors, and one
expects the coupling of the spins to be described by the Hamiltonian of a
Heisenberg spin chain:
\begin{equation}
  \label{eq:Heisenberg}
  H_\sigma = \sum_l J\,{\bm S}_{l}\cdot{\bm S}_{l+1}.
\end{equation}
Since the ground state of a system of interacting fermions in one
dimension cannot be spin-polarized~\cite{mattis}, the exchange must have
antiferromagnetic sign, $J>0$.

A careful evaluation of the exchange constant $J$ presents a challenging
problem of many-body physics, which is beyond the scope of this paper.  A
crude estimate of $J$ can be obtained by calculating the amplitude of
tunneling through the Coulomb barrier $e^2/\varepsilon r$ separating two
neighboring electrons in the WKB approximation.  Placing the turning
points at $r=\pm n^{-1}$ and using the reduced mass $m/2$, one finds
\begin{equation}
  \label{eq:J}
  J \sim E_F \exp\left(-\frac{\eta}{\sqrt{na_B}}\right)
\end{equation}
with the numerical coefficient $\eta=\pi$.  (Here $E_F\equiv \pi^2\hbar^2
n^2/8m$ is defined as the Fermi energy of a non-interacting electron gas
of density $n$.)  A more careful calculation will likely result in a
different value of $\eta$, but the exponential dependence of $J$ on
$na_B$ will remain.

The Hamiltonian of the Heisenberg spin chain (\ref{eq:Heisenberg}) can be
rewritten in terms of spinless fermion operators $a_l$ and $a_l^\dagger$
with the help of the Jordan-Wigner transformation
\begin{equation}
  \label{eq:Jordan-Wigner}
  S_l^z = a_l^\dagger a_l - \frac12,
  \quad
  S_l^x + iS_l^y 
      = a_l^\dagger\exp\left(i\pi\sum_{j=1}^{l-1} a_j^\dagger a_j\right).
\end{equation}
To study the low-energy properties of the spin chain, one can bosonize the
fermion operators $a_l$ and $a_l^\dagger$.  As a result the Hamiltonian
$H_\sigma$ takes the form (\ref{eq:H_sigma}), see, e.g.,
Ref.~\onlinecite{schulz}.  The velocity $u_\sigma$ of the spin excitations
is easily deduced from the spectrum \cite{fadeev} of the isotropic
Heisenberg model,
\begin{equation}
  \label{eq:u_sigma}
  u_\sigma = \frac{\pi J}{2\hbar n}.
\end{equation}
The sin-Gordon perturbation in Eq.~(\ref{eq:H_sigma}) is marginally
irrelevant, i.e., the coupling constant $g_{1\perp}$ scales to zero at low
energies.  At the same time the parameter $K_\sigma$ scales to 1, as
required by the SU(2) symmetry of the problem \cite{schulz}, \footnote{The
  conventional form of the bosonized Hamiltonian of a Heisenberg spin
  chain \cite{schulz} coincides with Eq.~(\ref{eq:H_sigma}) upon the
  transformation $\phi=\phi_\sigma/\sqrt2$, $\Pi=\sqrt2\,\Pi_\sigma$,
  $K=K_\sigma/2$.}.

It is important to stress that the bosonized form (\ref{eq:H_sigma}) of
$H_\sigma$ is only appropriate at low temperatures, $T\ll J$.  In the
following we will also be interested in the temperature dependence of the
conductance at $T\sim J$, and thus we will use the form
(\ref{eq:Heisenberg}).  Interestingly, the dynamics of spin and charge
modes are still completely separated, as the operators (\ref{eq:H_rho})
and (\ref{eq:Heisenberg}) commute.  We now show that this symmetry is
violated when the wire is connected to Fermi-liquid leads.

Following Ref.~\onlinecite{maslov}, we model the leads attached to the
quantum wire by two semi-infinite sections of non-interacting electron
gas.  To this end we assume that the 1D electron density $n(x)=n$ near
$x=0$, and gradually grows to a very high value $n_\infty\gg a_B^{-1}$ at
$x\to\pm\infty$.  Assuming that the length scale $L$ of the dependence
$n(x)$ is large compared to the distance between electrons, one can
neglect the backscattering caused by the inhomogeneity, and describe the
charge excitations by the bosonized Hamiltonian (\ref{eq:H_rho}) with
position-dependent parameters $u_\rho(x)$ and $K_\rho(x)$.  In addition,
the coupling constant $J$ in the Hamiltonian (\ref{eq:Heisenberg}) now
depends on $l$ due to the density dependence (\ref{eq:J}) of the exchange
interaction.  The exchange constant $J[l]$ is determined by the density
$n(x_l)$ at the position of the $l$-th electron.

Now let us consider the quantum wire in the regime when an electric
current $I=I_0\cos\omega t$ passes through it.  We will be interested in
the dc limit $\omega\to 0$, and can thus assume that all electrons move in
phase.  The charge transferred through any point in the wire is
$Q=I_0\omega^{-1} \sin\omega t$, so at moment $t$ the $l$-th electron
has shifted to the position $l+I_0(e\omega)^{-1} \sin\omega t$.  Thus the
Hamiltonian of the spin chain takes the form
\begin{equation}
  \label{eq:inhomogeneous}
  H_\sigma = \sum_l J[l+q(t)]\,
             {\bm S}_{l}\cdot{\bm S}_{l+1},
  \quad
  q(t) = \frac{I_0}{e\omega}\sin\omega t.
\end{equation}

One can view current $I=e\dot q$ as an excitation of the charge mode
$\phi_\rho$ and substitute the appropriate relation
\begin{equation}
  \label{eq:boundary_condition}
  q(t)=\frac{\sqrt2}{\pi}\phi_\rho(0,t)
\end{equation}
into the Hamiltonian (\ref{eq:inhomogeneous}).  Therefore the spin modes
are coupled to the charge ones and should affect the conductance of the
wire.

In the following it will be more convenient to treat the current
$I(t)=I_0\cos\omega t$ as an external parameter.  This approach
corresponds to the experiment with the wire connected to a current source.
We will evaluate the energy $W$ dissipated in the device in unit time in
the limit of small $I_0$ and $\omega$.  The dc resistance will be found
from the Joule heat law $W=\frac12 I_0^2 R$.  The advantage of this
approach is that the spin and charge modes are coupled only through the
external parameter $I(t)$, so $W$ will be a sum of two independent
contributions of the charge and spin modes.  Thus the resistance of the
wire is the sum of two terms $R=R_\rho+R_\sigma$, which can be evaluated
by considering the Hamiltonians $H_\rho$ and $H_\sigma$ separately.

The result $G=2e^2/h$ for the conductance of a quantum wire found in
Ref.~\onlinecite{maslov} amounts to the calculation of the resistance
$R_\rho$, as the spin modes were ignored.  Thus one expects to find
$R_\rho=h/2e^2$.  Let us outline the derivation of this result in the
approach where the current $I(t)$ through the wire is fixed.  Then the
charge subsystem is described by the Hamiltonian (\ref{eq:H_rho}) with the
boundary condition (\ref{eq:boundary_condition}).  It will be convenient
to transform the variables $\phi_\rho(x)\to \phi_\rho(x)+\pi q(t)/\sqrt2$,
i.e., to apply to the Hamiltonian a unitary transformation
\begin{equation}
  \label{eq:U}
  U=\exp\left(-i\frac{\pi q(t)}{\sqrt2}
              \int_{-\infty}^\infty \Pi_\rho(x)dx\right).
\end{equation}
Upon this transformation the boundary condition becomes time-independent,
$\phi_\rho(0)=0$, and the Hamiltonian transforms into $\tilde
H_\rho=U^\dagger H_\rho U -i\hbar U^\dagger \partial_t U$.  The second
term is a time-dependent perturbation that excites plasmons of very low
frequency $\omega$.  The wavelength of these plasmons is large,
$u_\rho/\omega\gg L$, and thus one can replace $u_\rho$ and $K_\rho$ by
their values at $x\to\infty$; in particular, $K_\rho(\infty)=1$, as
$n_\infty a_B\gg1$.  The transformed Hamiltonian $\tilde H_\rho$ is
conveniently presented in terms of the operators
\[
   b_k=\int_{-\infty}^\infty \theta(kx)\sin kx
      \left(\frac{\sqrt{|k|}}{\pi}\phi_\rho(x)
            +\frac{i}{\sqrt{|k|}}\Pi_\rho(x)
      \right)dx.
\]
destroying plasmons with frequency $\omega_k=u_\rho(\infty)|k|$.  Then
$\tilde H_\rho$ takes the form
\begin{equation}
  \label{eq:H_transformed}
  \tilde H_\rho=\hbar\int_{-\infty}^\infty\left(
           \omega_k b_k^\dagger b_k
         +i\frac{I(t)}{e}\frac{b_k-b_k^\dagger}{\sqrt{2|k|}}
          \right)dk. 
\end{equation}
At low frequency $\omega\ll T$ the time-dependent perturbation leads to
both emission and absorption of plasmons with $k=\pm
\omega/u_\rho(\infty)$.  The energy dissipated into plasmon excitations in
unit time is easily found by means of the Fermi golden rule, resulting in
$W=\frac12 I_0^2 (\pi\hbar/e^2)$.  Thus we conclude that $R_\rho=h/2e^2$.

To find $R_\sigma$ one has to perform a similar calculation with a more
complicated Hamiltonian (\ref{eq:inhomogeneous}).  Performing the
Jordan-Wigner transformation (\ref{eq:Jordan-Wigner}), we rewrite it as
\begin{eqnarray}
    H_\sigma &=& \frac12 \sum_l J[l+q(t)]\biggl[
               \left(a_l^\dagger a_{l+1} + a_{l+1}^\dagger a_{l}\right)
\nonumber\\
             && + 2\left(a_l^\dagger a_l + \frac12\right)
                       \left(a_{l+1}^\dagger a_{l+1} + \frac12\right)
                 \biggl].
  \label{eq:fermionic}
\end{eqnarray}
In the absence of the external magnetic field the average spin per site
$\langle S^z_l\rangle=0$.  In terms of the Hamiltonian
(\ref{eq:fermionic}) it means that the fermionic band is half-filled,
$\langle a_l^\dagger a_l\rangle = \frac12$.  The exchange $J[l]$ takes its
lowest value $J$ at the center of the wire and grows up to $J[\infty]\sim
E_F$ in the leads.

The Hamiltonian (\ref{eq:fermionic}) can be easily treated if one neglects
the interaction term in its second line.  (This corresponds to the XY
model of a spin chain, in which the $z$-component of exchange vanishes.)
Then the Hamiltonian describes a tight-binding model of non-interacting
fermions with the bandwidth $2J[l]$ varying gradually between the small
value $2J$ in the wire and a large value $\sim E_F$ in the leads.  The
fermions with energies $|\epsilon|<J$ pass through the constriction
without backscattering, whereas the ones with $|\epsilon|>J$ are
reflected.  The time dependence $J[l+q(t)]$ can be interpreted as a slow
movement of the constriction with respect to the Fermi gas, and leads to
the dissipation of energy.  For non-interacting fermions the calculation
of $W$ and the respective resistance is rather straightforward, and we
find
\begin{equation}
  \label{eq:XY}
  R_\sigma^{XY}= \frac{h}{2e^2} \frac{1}{e^{J/T}+1}.
\end{equation}
At $T\ll J$ this result is exponentially small, $R_\sigma^{XY}\propto
e^{-J/T}$, because at low temperature most of the quasiparticles pass the
constriction without scattering.  Only an exponentially small fraction of
the excitations are scattered at the constriction and acquire the energy
from the driving field.  The resistance saturates at $T/J\to\infty$, when
all the quasiparticles are backscattered.

The result (\ref{eq:XY}) remains qualitatively correct for the full model
(\ref{eq:fermionic}).  In particular, at $J\ll T$ the transport of
spin-fermions through the constriction is still suppressed, and $R_\sigma$
saturates.  In order to find $R_\sigma$ at $T/J\to\infty$, one can notice
that the bosonized model (\ref{eq:H_sigma}) is still applicable in the
leads, where $J\sim E_F\gg T$. Then the absence of spin transfer through
the constriction can be expressed as a hard-wall boundary condition upon
$\phi_\sigma$.

At $q(t)=0$ this boundary condition may be presented without loss of
generality as $\phi_\sigma(0)=0$.  At non-zero $q(t)$ the same condition
$\phi_\sigma=0$ is imposed at point $l=-q(t)$.  The half-filling condition
for the Hamiltonian (\ref{eq:fermionic}) means that $q(t)/2$ fermions have
passed through the wire at moment $t$.  In analogy with the calculation of
$R_\rho$, the boundary condition can be interpreted as application of a
fixed current $\dot q(t)/2$ of the spin-fermions.  The bosonization
procedure leading from Hamiltonian (\ref{eq:fermionic}) to
Eq.~(\ref{eq:H_sigma}) expresses the current of spin-fermions as
$\partial_t\phi_\sigma/\pi\sqrt2$.  Thus the appropriate boundary
condition for the bosonized spin Hamiltonian (\ref{eq:H_sigma}) is
$\frac{\sqrt2}{\pi} \phi_\sigma(0,t)=q(t)$.  Note, that this boundary
condition is equivalent to Eq.~(\ref{eq:boundary_condition}), and the
Hamiltonians (\ref{eq:H_rho}) and (\ref{eq:H_sigma}) coincide in the
leads, where $K_\rho=K_\sigma=1$ and $g_{1\perp}=0$.  Therefore one can
complete the evaluation of $R_\sigma$ by repeating the above calculation
of $R_\rho$, and we conclude that $R_\sigma=h/2e^2$.  The conductance of
the wire $(R_\rho+R_\sigma)^{-1}$ reduces to $e^2/h$.

It is worth mentioning that the same boundary condition for the
Hamiltonian (\ref{eq:H_sigma}) appears even at $T\lesssim J$ if a
sufficiently large magnetic field $B$ is applied.  Indeed, if the Zeeman
splitting is large compared to both $T$ and the coupling $J$, the
electrons in the wire are completely spin polarized, $\langle
S^z_l\rangle= -\frac12$.  In terms of the Hamiltonian (\ref{eq:fermionic})
this is interpreted as lowering of the chemical potential below the bottom
of the fermionic band in the wire, so that $\langle a_l^\dagger
a_l\rangle=0$.  Thus the two leads are now separated by a barrier created
by the narrowing fermionic band in the wire.  The barrier is centered at
$l=-q(t)$, and, upon bosonization, imposes the same boundary condition
upon $\phi_\sigma$ as in the case of $B=0$ and $T>J$.  We therefore
conclude that in a polarizing magnetic field $R_\sigma=h/2e^2$, and, as
expected, the total conductance $G=e^2/h$.

At low temperature $T\ll J$ the model (\ref{eq:fermionic}) to first
approximation can be bosonized to the form (\ref{eq:H_sigma}) with
position-dependent parameters $u_\sigma$, $K_\sigma$ and $g_{1\perp}$.  At
$T\to0$ we have $K_\sigma=1$ and $g_{1\perp}=0$, i.e., the Hamiltonian
(\ref{eq:H_sigma}) can be viewed as the bosonized version of the XY model.
Then we conclude from Eq.~(\ref{eq:XY}) that at zero temperature
$R_\sigma=0$, and the conductance of the wire remains $2e^2/h$.

At finite length of the wire $L$ the parameters of the Hamiltonian
(\ref{eq:H_sigma}) do not reach their limiting values $K_\sigma=1$,
$g_{1\perp}=0$.  However, the resulting corrections to $R_\sigma$ remain
small in $1/nL$ even at $T\sim J$.  A more interesting correction appears
due to the fact that the bandwidth of the Hamiltonian (\ref{eq:fermionic})
is finite, which is not accounted for accurately by the bosonization
procedure leading to Eq.~(\ref{eq:H_sigma}).  If the wire is long,
$nL\gg1$, the spin chain (\ref{eq:inhomogeneous}) is nearly uniform at
each point $l$.  Its low energy excitations are conveniently classified
\cite{fadeev} in terms of spinon quasiparticles with spectrum $\epsilon =
\frac{\pi}{2} J[l] \sin p$, where $p$ is the wavevector in the lattice
model.  At low temperature $T\ll J$ typical excitations have energies
$\epsilon\sim T$ small compared to the bandwidth in the center of the
constriction.  These spinons pass through the wire without scattering and
do not interact with the driving field.  On the other hand, a small
fraction of the spinons have energies exceeding $\pi J/2$.  These
excitations are not supported by the spin chain at the center of the wire,
and thus are reflected by the constriction.  Similar to the case of the XY
model, the reflected particles acquire energy from the driving field and
contribute to the dissipation.  The density of such spinons is
exponentially small, and we get
\begin{equation}
  \label{eq:Rspinon}
  R_\sigma \sim R_0 \exp\left(-\frac{\pi J}{2T}\right).
\end{equation}
This treatment does not enable one to evaluate $R_0$.  The analogy with
the XY model suggests $R_0\sim h/e^2$.

Experimentally, corrections to the quantized conductance showing activated
temperature dependence consistent with (\ref{eq:Rspinon}) have been
reported in short wires \cite{kristensen}.  At higher temperatures the
conductance tends to saturate at $0.7\times(2e^2/h)$ instead of $e^2/h$.
Quantization of conductance at $e^2/h$ in the absence of magnetic field
was reported in longer wires at low electron density \cite{reilly-thomas}.

To summarize, in the regime of strong interactions, $na_B\ll1$, the
electrons in a quantum wire form a Wigner crystal with exponentially small
antiferromagnetic spin exchange $J$.  At $T\ll J$ the conductance remains
$2e^2/h$ up to exponentially small corrections, Eq.~(\ref{eq:Rspinon}).
At $J\ll T$ the conductance drops to $e^2/h$.  Remarkably, this result
does not assume spontaneous spin polarization in the wire
\cite{reilly2}.

\begin{acknowledgments}
  The author acknowledges helpful discussions with A.~V.  Andreev, A. M.
  Finkel'stein, A.~I.  Larkin, R. de Picciotto, and P.~B. Wiegmann and the
  hospitality of Bell Laboratories where most of this work was carried
  out.  This work was supported by the Packard Foundation and by NSF Grant
  DMR-0214149.
\end{acknowledgments}
\vspace{-\baselineskip}

\end{document}